\documentclass[10pt,aps,prx,preprintnumbers,twocolumn,notitlepage,
floatfix,unsortedaddress,longbibliography]{revtex4-1}

\usepackage{amsmath, amsthm, amssymb, amsfonts, amsbsy, mathrsfs}

\usepackage{xcolor}
\usepackage{calligra,bm}
\usepackage{stmaryrd}
\usepackage{wasysym}
\usepackage{yfonts}

\newcommand{\ls}{{%
\fontencoding{TS1}\fontfamily{lmr}\selectfont\char115}}

\usepackage{graphicx}
\graphicspath{{./images/}}

\usepackage[hidelinks,bookmarks=true]{hyperref}
\hypersetup{pdfstartview=FitH,pdfhighlight=/O,colorlinks=false}

\begin{document}
\preprint{YITP-23-134,  IPMU23-0039}

\title{Singularity at the demise of a black hole}

\author{Justin C. Feng}
\email{jcfeng@ntu.edu.tw}
\affiliation{Leung Center for Cosmology and Particle Astrophysics,
National Taiwan University, No.1 Sec.4, Roosevelt Rd., Taipei 10617,
Taiwan, Republic of China}

\author{Shinji Mukohyama}
\affiliation{Center for Gravitational Physics and Quantum Information
(CGPQI), Yukawa Institute for Theoretical Physics (YITP),\\ Kyoto
University, Kyoto 606-8502, Japan} \affiliation{Kavli Institute for the
Physics and Mathematics of the Universe (WPI), \\The University of Tokyo
Institutes for Advanced Study, \\The University of Tokyo, Kashiwa, Chiba
277-8583, Japan}

\author{Sante Carloni}
\affiliation{Institute of Theoretical Physics, Faculty of Mathematics
and Physics, Charles University, Prague, V Hole\v sovi\v ck\'ach 2, 180
00 Prague 8, Czech Republic}
\affiliation{DIME, Università di Genova, Via all'Opera Pia 15, 16145
Genova, Italy}
\affiliation{INFN Sezione di Genova, Via Dodecaneso 33, 16146 Genova,
Italy}

%-----------------------------------------------------------------------
%-----------------------------------
%-----------------
%--------
%---
%-
%
%
%-
%---
%--------
%-----------------
%-----------------------------------
%-----------------------------------------------------------------------

%=======================================================================
%-----------------------------------------------------------------------
%
%		ABSTRACT
%
%-----------------------------------------------------------------------
%=======================================================================
\begin{abstract}
    We consider a class of quasiregular singularities characterized by
    points possessing two future-directed light cones and two
    past-directed light cones. Such singularities appear in the $1+1$
    trousers spacetime and the Deutsch-Politzer spacetime. We argue that
    these singularities are relevant for describing the end point of an
    evaporating black hole, and show that a class of emergent Lorentz
    signature theories can provide a microscopic description for these
    singularities.
\end{abstract}

%-----------------------------------------------------------------------
%-----------------------------------
%-----------------
%--------
%---
%-
%
%
%-
%---
%--------
%-----------------
%-----------------------------------
%-----------------------------------------------------------------------

\maketitle

%-----------------------------------------------------------------------
%-----------------------------------
%-----------------
%--------
%---
%-
%
%
%-
%---
%--------
%-----------------
%-----------------------------------
%-----------------------------------------------------------------------

%=======================================================================
%-----------------------------------------------------------------------
%
%		Introduction
%
%-----------------------------------------------------------------------
%=======================================================================

\section{Introduction}
About half a century ago, Stephen Hawking considered quantum effects in
matter fields on a black hole spacetime, concluding that black holes
radiate \cite{Hawking:1974rv,Hawking:1975vcx}. If black holes radiate
and lose mass (i.e. they evaporate), one is led to the question: what is
the outcome of black hole evaporation? At least three possible answers
have been proposed in the literature \cite{Chen:2014jwq}: 1) a black
hole remnant, 2) a naked singularity
\cite{DeWitt:1975ys,Kodama:1979,Penrose:1980ge,Penrose:1999vj}, and 3) a
complete evaporation of the black hole to a region of flat spacetime
(see, for instance \cite{Hayward:2005gi}). The remnant scenario has been
studied extensively in literature (see the review \cite{Chen:2014jwq}
and references therein). In this article, we focus on the latter two,
which need not be mutually exclusive---in particular, the nature of the
naked singularity, which we show to be closely related to so-called
``crotch singularity'' of the trousers spacetime \cite{Anderson:1986ww}
in the $1+1$ dimensional setting, and a scenario in which the
singularity is regularized by a candidate ultraviolet (UV) completion
for quantum gravity.

Quasiregular singularities \cite{Ellis:1977pj} are singularities on
codimension two surfaces, which may be regarded as generalizations of
conical singularities that, in addition to angle surpluses and deficits,
include surpluses and deficits in boosts, as well as dislocations
\cite{Krasnikov:2006zx} and additional light cones (cf. Fig. 3 of
\cite{Ellis:1977pj}); a further generalization can be found in
\cite{bruno1987quasi}. The latter constitutes a radical change to the
causal structure of spacetime on a surface of codimension two---such
points are termed {\it{local causal discontinuities}}
\cite{Hawking:1974ar}, in which the local past and future light cones of
an instantaneous observer fail to deform continuously as the observer is
displaced in the manifold; the structure of these types of causal
discontinuities have been studied in
\cite{Dowker:1997hj,Dowker:1999wu,Borde:1999md,Garcia-Heveling:2022fkf}.
The best-known example of such a singularity is the crotch singularity
of the $1+1$ trousers spacetime \cite{Anderson:1986ww} featuring two
future-directed light cones and two past-directed light cones. The $1+1$
trousers spacetime provides a simple model for studying the effects of
topology change on quantum fields and in quantum gravity
\cite{Anderson:1986ww,manogue1988trousers,Ishibashi:2002ac,
Barvinsky:2012qm,Krasnikov:2016ypz,Louko:1995jw,Dowker:2002hm,
Bauer:2006pg}. Another example is the singularity present in the
Deutsch-Politzer spacetime \cite{Deutsch:1991nm,Politzer:1992zm} (and
the formally homeomorphic teleporter spacetimes \cite{Feng:2021pfz}),
which have a similar causal structure to that of the trousers spacetime
at the singular surfaces. The reader can find a discussion of the
geometric and topological properties of the Deutsch-Politzer spacetime
in \cite{Gibbons:1992he,Chamblin:1994ed,Yurtsever:1994bs}.

Though the study of quasiregular singularities is primarily motivated by
topology change in quantum gravity, and a case for studying such
singularities was made in \cite{Krasnikov:2009xt}, interest in the topic
has mainly focused on the implications of similar causal structures
within the causal set approach to quantum gravity
\cite{Dowker:2002hm,Benincasa:2010as,Buck:2014inj,Buck:2016ehk,
      Jones:2021orh}.
Here, we will attempt to bring a sense of urgency to the study of
quasiregular singularities by making the case that if no mechanism stops
the continued evaporation of a black hole (such as that arising from the
generalized uncertainty principle \cite{Adler:2001vs}, for instance)
then one would expect the disappearance of a black hole event horizon at
the end point of black hole evaporation. The end state of the black hole
will contain a local causal discontinuity closely related to those of
quasiregular singularities---in fact, we demonstrate explicitly that in
the $1+1$ case, the end point of a black hole that evaporates completely
contains a quasiregular singularity of the same type as that of crotch
singularity in the trousers spacetime, provided that the Schwarzschild
singularity is regularized so that the resulting manifold is Lorentzian
everywhere except for the singular point.

Of course, one might expect a microscopic description of causal
discontinuities and quasiregular singularities to require a UV
completion for quantum gravity---indeed, as the reader can see in the
papers referenced thus far, such a possibility has been explored in some
detail in the causal set approach to quantum gravity. Here, we will
consider a model motivated by a more conservative candidate for a UV
completion. The candidate in question is a quadratic curvature,
shift-symmetric Euclidean scalar-tensor theory proposed a decade ago in
\cite{Mukohyama:2013gra}. This theory was shown to be renormalizable
\cite{Muneyuki:2013aba} and evades the Ostrogradsky instability (one
only requires the action to be bounded below since the theory is
fundamentally Euclidean, as discussed in
\cite{Mukohyama:2013gra}). In the long-distance limit, one may recover
an effective Lorentz signature metric with an appropriate coupling to
the matter degrees of freedom (see
\cite{Mukohyama:2013ew,Kehayias:2014uta} for a detailed discussion of
this mechanism). In this theory, a scalar field with a saddle profile
can mimic the long-distance features of a quasiregular singularity, and
we will show that there exists an exact saddle solution for a particular
set of parameter choices, and an approximate saddle solution in the
vicinity of a point for generic choices of parameters.

This article is organized as follows. In Sec. \ref{Sec:SCDS}, we discuss
the singularities of the trousers spacetime, the relationship between
the trousers spacetime and an evaporating $1+1$ black hole. In Sec.
\ref{Sec:ELSTs}, we review the theory of \cite{Mukohyama:2013gra} and
obtain solutions that regularize the quasiregular singularity at the end
point of an evaporating black hole (assuming an either Lorentzian or
Euclidean regularization of the Schwarzschild singularity), and in Sec.
\ref{Sec:Disc}, we discuss implications and future directions.

%=======================================================================

%-----------------------------------------------------------------------
%-----------------------------------
%-----------------
%--------
%---
%-
%
%
%-
%---
%--------
%-----------------
%-----------------------------------
%-----------------------------------------------------------------------

%=======================================================================
%-----------------------------------------------------------------------
%
%   Saddlelike singularities and black hole evaporation
%
%-----------------------------------------------------------------------
%=======================================================================
\section{Saddlelike singularities and black hole evaporation}
\label{Sec:SCDS}

%-----------------------------------------------------------------------
%   Construction
%-----------------------------------------------------------------------
\subsection{Construction}
\begin{figure}[!ht]
    \includegraphics[width=1.0\columnwidth]{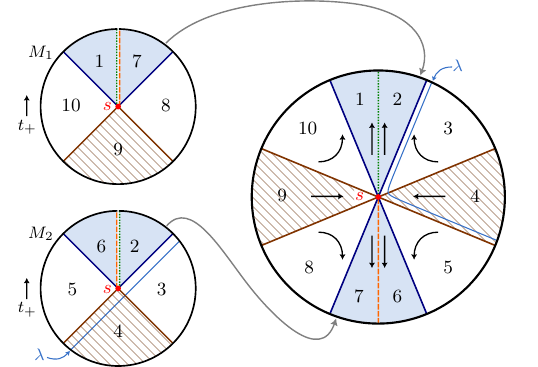}
    \caption{Cut-and-paste procedure for the construction of an SCDS. On
    the left are illustrations of two regions of $1+1$ Minkowski
    spacetime $M_1$ and $M_2$. A caricature of the result is on the
    right (in which angles are {\em{not}} preserved---null directions on
    the right are {\it not} $45^\circ$ lines, as illustrated by the
    behavior of the null line $\lambda$). The solid black arrows
    indicate the positive time direction in each region.}
    \label{fig:QRSGluing}
\end{figure}

The simplest causally discontinuous quasiregular singularity can be
constructed by performing a simple cut-and-paste
procedure \footnote{Similar cut-and-paste procedures can be used
to construct wormholes supported by cosmic strings, as described in
\cite{Visser1989,*Visser1995}.} in two regions of $1+1$ Minkowski space,
which we call $M_1$ and $M_2$ [each equipped with coordinates $(x,t)$],
as indicated in Fig \ref{fig:QRSGluing}; one can find similar diagrams
in \cite{Ellis:1977pj} and \cite{harris1990causal} [cf. Fig. 4e in the
former and Figs. 17 and 18 in the latter]. In the regions $M_1$ and
$M_2$ on the left side of Fig. \ref{fig:QRSGluing}, one performs a cut
along the positive time axis indicated by the dotted lines, and are
glued so that region $1$ is glued to region $2$ and region $6$ is glued
to region $7$. Regions $1,2$ and $6,7$ correspond to future light cones
of $s$, and regions $9$ and $4$ correspond to past light cones of $s$. A
caricature of the result is on the right of Fig. \ref{fig:QRSGluing}. As
one can infer, the origin $s$ is causally discontinuous, as points in
the neighborhood of $s$ other than $s$ itself have one future-directed
light cone and one past-directed light cone, while the point $s$ has two
future-directed and two past-directed light cones. From this point, we
will refer to a naked singularity with such a causal structure as a
{\em{saddlelike causally discontinuous singularity}} (SCDS); in
particular, we define a SCDS to be a singular surface $s$ with a
spacelike quasiregular singularity (as defined in \cite{Ellis:1977pj})
such that each point on $s$ has two opposed future-directed light cones,
and two opposed past-directed light cones. The name refers the fact that
the surfaces of constant $t$ behave as contours of a saddle, as one
might infer from the direction of increasing time indicated on the 
right-hand side of Fig. \ref{fig:QRSGluing}.

%-----------------------------------------------------------------------
%   Trousers spacetime and black hole evaporation
%-----------------------------------------------------------------------
\subsection{Trousers spacetime and black hole evaporation}

\begin{figure*}[!ht]
    \includegraphics[width=1.0\textwidth]{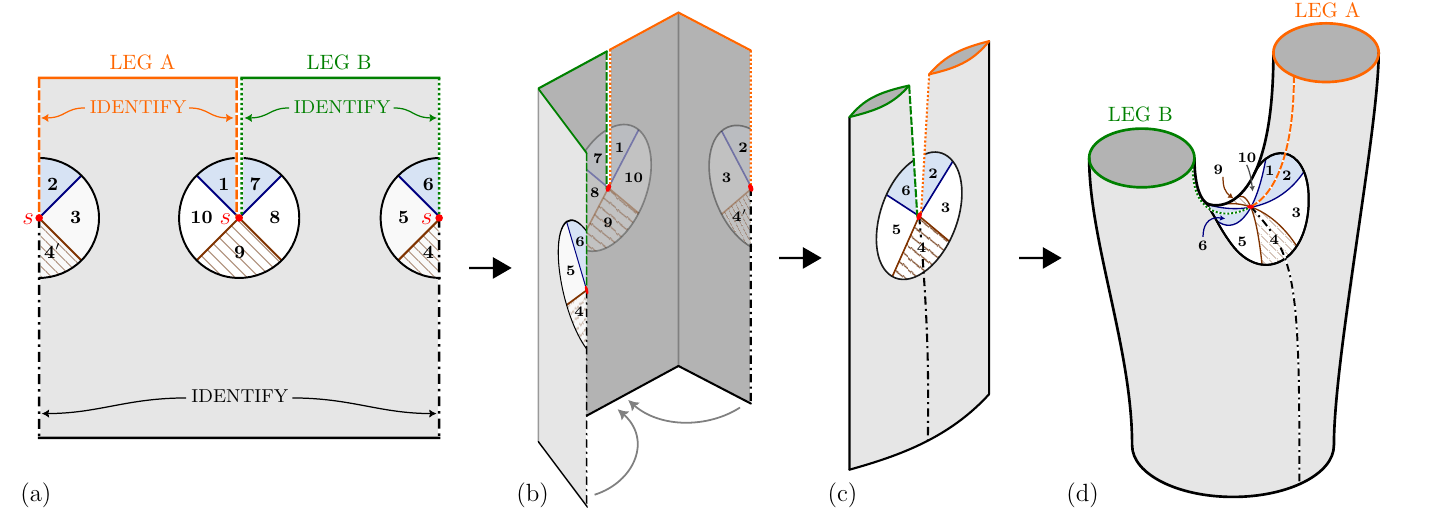}
    \caption{(a) depicts a flat $1+1$ trousers spacetime and the
    neighborhood $N_s$ of the crotch singularity $s$, with the time
    direction oriented vertically. (b) and (c), respectively, illustrate 
    a folding and gluing procedure to construct a paper model of a $1+1$
    trousers spacetime [observe that regions $4^\prime$ and $4$ of (a)
    and (b) have been combined into a single region $4$ here to
    facilitate comparison with Fig. \ref{fig:QRSGluing}]. (d)
    illustrates a homeomorphic deformation of the $1+1$ trousers
    spacetime to illustrate the neighborhood $N_s$ more clearly. In all
    diagrams, the solid shaded regions of $N_s$ indicate the future
    light cones of $s$, and the crosshatched regions indicate the past
    light cones of $s$.}
    \label{fig:TrousersQRS}
\end{figure*}

The $1+1$ trousers spacetime provides a simple example of topology
change and contains a ``crotch'' singularity $s$ featuring two distinct
future-directed light cones and two distinct past-directed light cones,
making $s$ an SCDS. Figure \ref{fig:TrousersQRS} illustrates an
example of a trousers spacetime, and the SCDS $s$. Figure
\ref{fig:TrousersQRS} (a) through \ref{fig:TrousersQRS} (c) illustrates 
how one can construct a trousers spacetime from appropriate 
identifications in $1+1$ Minkowski spacetime \cite{Anderson:1986ww}, 
and Fig. \ref{fig:TrousersQRS} (d) is a homeomorphic deformation 
included to clearly illustrate the neighborhood of the singular point 
(cf. Fig. 3 of \cite{Buck:2016ehk}).

\begin{figure*}[]
    \includegraphics[width=0.90\textwidth]{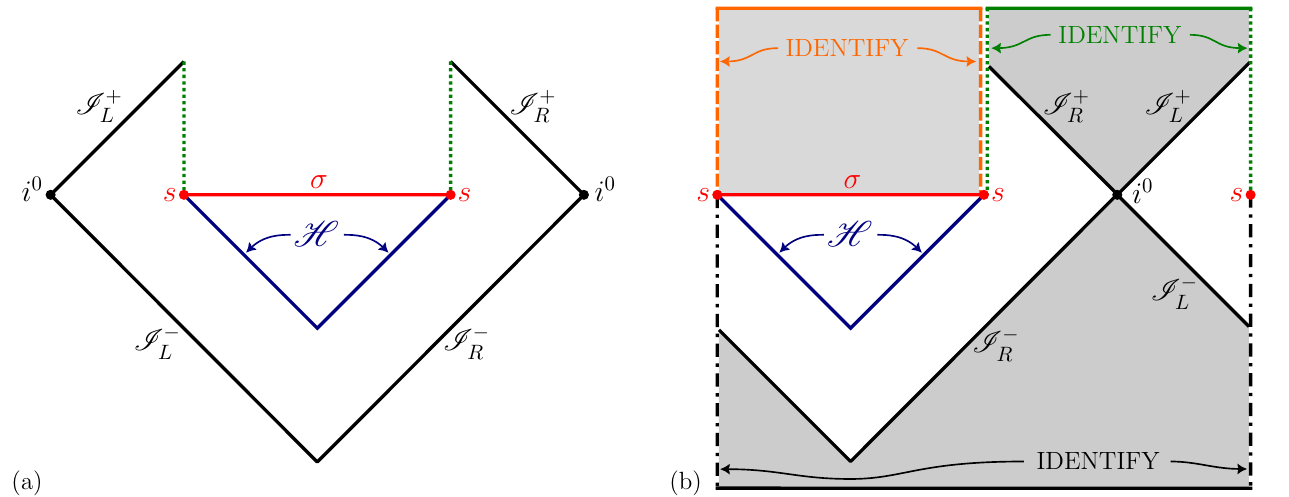}
    \caption{A full conformal diagram describing the formation and
    evaporation of a $1+1$ Schwarzschild-like black hole (a),
    accompanied by an illustration demonstrating a conformal embedding
    of the $1+1$ evaporating black hole in a region of the $1+1$
    trousers spacetime (b). In the $1+1$ case, the respective future and
    past null infinity $\mathscr{I}^+$ and $\mathscr{I}^-$ are split
    into two distinct lines distinguished by the subscripts $L$ and $R$.
    The lines labeled $\mathscr{H}$ represent the horizon and $\sigma$
    represents the classical singularity, which we suppose is
    regularized by quantum gravity effects. In (b), the dotted lines are
    identified, as are the points labeled $s$ and $i^0$, which are the
    evaporation point and spatial infinity, respectively.}
    \label{fig:BHevapTrousers}
\end{figure*}

\begin{figure*}[]
    \includegraphics[width=0.91\textwidth]{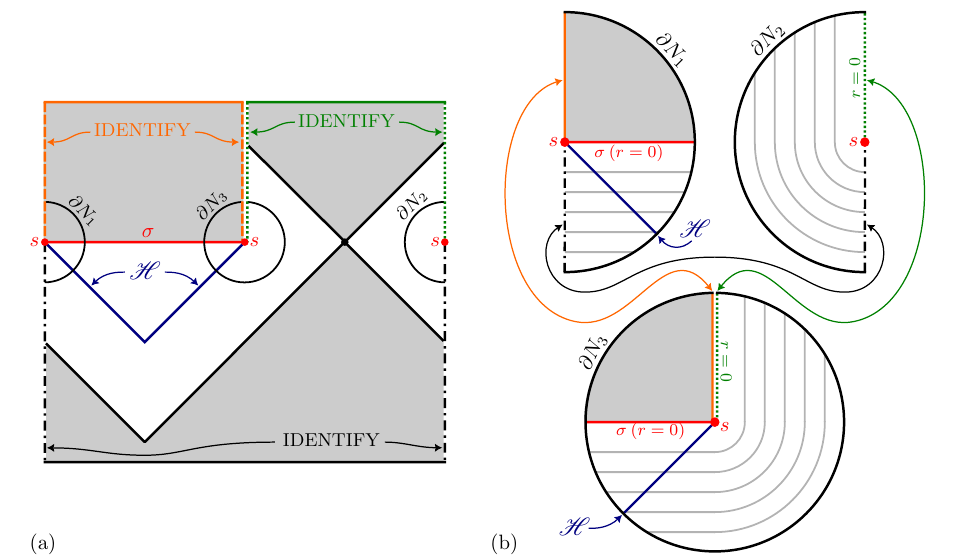}
    \caption{Neighborhood of a $d+1$ dimensional black hole end point
    $s$ on a surface $\Sigma_\Omega$. In (a), the neighborhood of $s$
    in the conformal diagram [cf. Fig. \ref{fig:BHevapTrousers} (b)],
    is the region enclosed by the boundaries $\partial N_{i}$,
    $i\in\{1,2,3\}$. (b) is a detailed diagram of the neighborhood
    (identifications indicated by arrows) with the gray contours
    describing surfaces of constant areal radius $r>0$.}
    \label{fig:RContours}
\end{figure*}

We offer here a physical motivation for considering the $1+1$ trousers
spacetime and its SCDS $s$ in detail. In particular, we point to the
observation that the full conformal diagram for a $1+1$ evaporating
Schwarzschild-like black hole can be pasted into the region indicated in
Fig. \ref{fig:BHevapTrousers} (b) on the flat trousers
spacetime\footnote{One might imagine that with regard to conformal
embeddings, the flat trousers spacetime is to the $1+1$ evaporating
black hole what the Einstein static cylinder is to Minkowski spacetime
(cf. Fig. 14 of \cite{HawkingEllis}).}, demonstrating that up to the
singularities, such a $1+1$ black hole is conformal to a subset of the
flat trousers spacetime. One finds that the singular point $s$ coincides
with the SCDS of a trousers spacetime. This is a rather suggestive
result, as one might imagine that even if one has in hand a $1+1$
Lorentzian quantum gravity theory that regularizes the usual classical
Schwarzschild singularity $\sigma$, the external causal structure of an
evaporating black hole requires the presence of an SCDS at $s$. If one
demands a regularization of the Schwarzschild singularity $\sigma$,
Fig. \ref{fig:BHevapTrousers} (b) suggests that upon regularization, the
Schwarzschild singularity will be replaced by a bounce and formation of
a new universe \footnote{Such a possibility has been suggested before,
for instance, in \cite{Sorkin:1989ea,Dowker:2002hm}. If one considers
the $d+1$ setting in spherical symmetry, a new universe resulting from
the regularization of the Schwarzschild singularity will be highly
anisotropic, having infinite extent in one direction and closed in the
remaining directions.} and from the results in
\cite{Anderson:1986ww,manogue1988trousers}, one might also expect some
rather extreme behavior (a ``thunderbolt''
\cite{penrose1978singularities,Hawking:1992ti,Penrose:1999vj}) arising
from the interaction between quantum fields and the SCDS at $s$, though
it has been argued that this problem can be avoided
\cite{Krasnikov:2016ypz,Schulz:2022vvp}. Of course, one can avoid these
issues outright by forbidding topology change, fixing the topology of
spacetime from the outset by fiat, and in such an approach, one must
either accept that the end state of a black hole is a remnant or discard
the notion of a global black hole event horizon altogether (which would
likely require an effective violation of energy conditions to avoid
singularities); though these are interesting scenarios, we shall not
consider them further in this article.

The reader might wonder about the extension of the aforementioned
discussion to the evaporation of a spherically symmetric $d+1$
dimensional black hole.
One might proceed in the $d+1$ dimensional case by reinterpreting the
conformal diagram in Fig. \ref{fig:BHevapTrousers} (a) as representing a
surface $\Sigma_\Omega$ of codimension two with two sets of antipodally
related angular values---that is, points reflected about the vertical
axis of symmetry represent antipodal points. The evaporation point $s$
then forms an SCDS in the surface $\Sigma_\Omega$ as one might infer
from Fig. \ref{fig:RContours} (a). Of course, the reader
may wish to gain a more complete understanding of the nature of the SCDS
at the evaporation end point of a spherically symmetric black hole and
the properties of its neighborhood. Though a full discussion is well
beyond the scope of this article, we offer a first step in that
direction. In particular, one may think of the areal radius $r$ as
defining a scalar field $r(x)$ [$x$ being a point on the manifold],
then consider possible contours of constant
$r>0$ in the vicinity of an SCDS in $\Sigma_\Omega$. Points in Fig.
\ref{fig:RContours} represent $d-1$ dimensional spheres, and Fig.
\ref{fig:RContours} (b) illustrates possible constant $r>0$ contours in
a neighborhood $N_s$ of the evaporation point $s$ for a $d+1$
dimensional spherically symmetric evaporating black hole. If $\sigma$
is regularized, it may no longer correspond to an $r=0$ surface, and
the green dotted $r=0$ contour represents a ``valley'' in the scalar
field profile $r(x)$ that ends at the point $s$.

\begin{figure}[]
    \includegraphics[width=1.0\columnwidth]{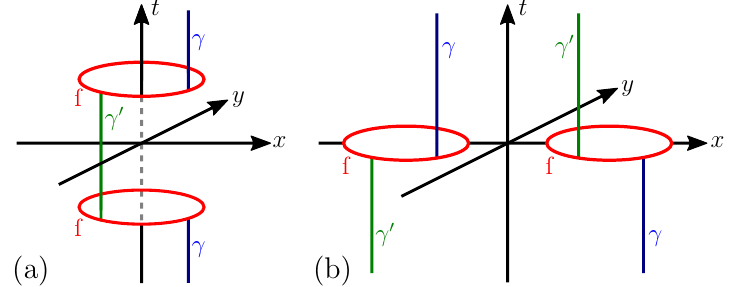}
    \caption{(a) is an illustration of a $2+1$ Deutsch-Politzer
    spacetime, and (b) is an illustration of a $2+1$ teleporter
    spacetime. In each case, the spacetimes are assumed to be flat
    everywhere except the surface {\ls}, which contain a SCDS. Both
    $\gamma$ and $\gamma^\prime$ are timelike curves [note that
    $\gamma^\prime$ in (a) is a closed timelike curve].
}
    \label{fig:DPTeleporter}
\end{figure}

%-----------------------------------------------------------------------
%   Time machines and teleporters
%-----------------------------------------------------------------------
\subsection{Time machines and teleporters}

Before proceeding, it is perhaps appropriate to remark on the
implications of SCDSs, if they are permitted to exist. One would be the
formation of Deutsch-Politzer spacetimes containing closed timelike
curves \cite{Deutsch:1991nm,Politzer:1992zm}, and another would be
teleporter spacetimes which feature dislocations that can
instantaneously (that is, along a spacelike hypersurface) ``transport''
timelike geodesics across large distances \cite{Feng:2021pfz}. One can
easily construct such spacetimes from a cut-and-paste procedure in
Minkowski spacetime. An illustration in the $2+1$ case is provided in
Fig. \ref{fig:DPTeleporter}. The cut-and-paste procedure is performed
along circular disks on constant $t$ planes such that the curve $\gamma$
passes through the disks in the manner indicated in Fig.
\ref{fig:DPTeleporter}. More generally, the gluing in each case is
performed to obtain a spacetime with topology
$\mathbb{S}^2\times\mathbb{S}^1-\{0\}$ (which generalizes to
$\mathbb{S}^{d}\times\mathbb{S}^1-\{0\}$ in the $d+1$ dimensional case);
the difference between the Deutsch-Politzer and teleporter spacetimes
being the presence of closed timelike curves in the former. Teleporter
spacetimes are defined so that all pairs of points on the surface {\ls}
are spacelike separated. Since such structures have not yet been
observed in nature, a theory which admits SCDSs should provide some
mechanism to suppress these types of structures at macroscopic scales.

%=======================================================================

%-----------------------------------------------------------------------
%-----------------------------------
%-----------------
%--------
%---
%-
%
%
%-
%---
%--------
%-----------------
%-----------------------------------
%-----------------------------------------------------------------------

%=======================================================================
%-----------------------------------------------------------------------
%
%   Regularized SCDS in emergent Lorentz signature theories
%
%-----------------------------------------------------------------------
%=======================================================================
\section{Regularized SCDS in emergent Lorentz signature theories}
\label{Sec:ELSTs}

Up to this point, we have considered the nature of SCDS at the end point
of an evaporating black hole. One might expect SCDSs to be ultimately
described by some UV completion for quantum gravity or some appropriate
modified gravity theory. In this section, we will consider one such
theory \cite{Mukohyama:2013gra}, showing that it yields a regularization
for SCDS even at the classical level.

%-----------------------------------------------------------------------
%   Emergent Lorentz signature theories
%-----------------------------------------------------------------------
\subsection{Emergent Lorentz signature theories}
We consider a class of theories described in \cite{Mukohyama:2013ew},
which we will henceforth refer to as {\it emergent Lorentz signature
theories} (ELSTs). ELSTs are scalar-tensor theories constructed from a
Euclidean signature metric $g_{ab}$ and a scalar $\varphi$ so that one
has an effective metric ${\bf{g}}_{ab}$ (and inverse
$\bar{\bf{g}}^{ab}$) of the form,
\begin{equation} \label{Eq:EffectiveMetric}
    {\bf{g}}_{ab} = g_{ab} -
                        \frac{\nabla_a \varphi \nabla_b\varphi}
                        {X_C},
                        \qquad
    \bar{\bf{g}}^{ab} = g^{ab} -
                        \frac{\nabla^a \varphi \nabla^b \varphi}
                        {X-X_C},
\end{equation}
where $X_C$ is a positive constant that appears in the matter couplings,
$X:=\nabla^a \varphi \nabla_a \varphi$, and we employ the convention
that indices are raised and lowered with the Euclidean signature metric
$g_{ab}$. The proposal in ELSTs is that in the long-distance limit,
matter couples to the metric $g_{ab}$ and scalar $\varphi$ degrees of
freedom exclusively through the effective metric ${\bf{g}}_{ab}$, which
has a Lorentzian signature for sufficiently large scalar field
gradients \footnote{A couple of remarks concerning matter couplings are
in order. An interesting question for future investigation concerns the
mechanism for ensuring that all matter degrees of freedom couple to the
same effective metric ${\bf{g}}_{ab}$. The behavior of matter in the
vicinity of the singular surfaces at which ${\bf{g}}_{ab}$ becomes
degenerate is an issue only if matter couples exclusively to
${\bf{g}}_{ab}$, but in general higher derivative terms coupled to
$g_{ab}$ will dominate at short distances, but are suppressed at long
distances.}. In this manner, Lorentz signature emerges through matter
couplings, as described in \cite{Mukohyama:2013ew,Kehayias:2014uta}.

In \cite{Mukohyama:2013gra}, an ELST is proposed on a four-dimensional
manifold $M$ in which the action has the following form:
\begin{equation} \label{Eq:ELSTAction}
    \begin{aligned}
    S = \int_M d^4x \sqrt{|g|} L,  \qquad\qquad
    L := L_0+L_2+L_4,
    \end{aligned}
\end{equation}
where (defining $\varphi_a:=\nabla_a\varphi$,
$\varphi_{ab}:=\nabla_a\nabla_b\varphi$)
\begin{equation} \label{Eq:ELSTLagrangian}
    \begin{aligned}
    L_0 &:= c_{11}, \\
    L_2 &:= c_{9} R + c_{10} X , \\
    L_4 &:=
        c_1 R^2 + c_2 R_{ab} R^{ab} + c_3 R_{abcd} R^{abcd} + c_4 X R\\
        & \quad \>
        + c_5 R^{ab}\varphi_a \varphi_b + c_6 {\rm X}^2
        + c_7 (\Box^2 \varphi)^2+ c_8 \varphi_{ab} \varphi^{ab} .
    \end{aligned}
\end{equation}
The couplings $c_I$ ($I\in\{1,...,11\}$) can be chosen so that the
action is bounded below.
The action is constructed so that it is shift symmetric and invariant
under $\varphi \rightarrow -\varphi$. Higher derivatives of the scalar
$\varphi$ and quadratic curvature terms are introduced so that the
theory is power-counting renormalizable \cite{Stelle:1976gc}; it was
later shown in \cite{Muneyuki:2013aba} that the theory defined by $S$ is
indeed perturbatively (super-) renormalizable in three and four 
dimensions. In the long-distance limit, the theory reduces to a 
Lorentzian scalar-tensor theory \cite{Mukohyama:2013gra}. The 
renormalizability of the resulting theory and the boundedness of the 
action below (since the theory is fundamentally Euclidean the latter 
allows it to evade the Ostrogradsky instability and its associated 
ghosts, a pathology that would plague a Lorentzian higher derivative 
theory \cite{Ostrogradsky1850,Woodard:2015zca}) makes it a potential
candidate as a UV completion for quantum gravity. Moreover, the
renormalizability of the theory makes it useful as an effective field
theory even if the UV completion takes a radically different form, so
long as it retains symmetries that reduce to the shift symmetry,
diffeomorphism invariance and local $SO(4)$ invariance seen in the
action described by Eqs. \eqref{Eq:ELSTAction} and
\eqref{Eq:ELSTLagrangian}. Of course, while this discussion is motivated
by quantum gravitational considerations, for simplicity, we shall focus
on {\em classical} solutions of the theory.

Since the action $S$ is shift symmetric in $\varphi$, the field equation
corresponding to variations in $\varphi$ has the form $\nabla_a
(J^a-\Phi^a)=0$, where $\Phi^a$ is the contribution from the matter
degrees of freedom and the current $J^a$ is
\begin{equation} \label{Eq:ELSTFieldEquationsScalarCur}
    J^a =   (c_{7} + c_{8}) \Box\varphi^{a}
            - (c_{5} + c_{7}) R^{ab} \varphi_{b}
            - \varphi^{a}
            (c_{10} + c_{4} R + 2 c_{6} X).
\end{equation}
The field equation corresponding to variations in $g_{ab}$ takes the
form,
\begin{equation} \label{Eq:ELSTFieldEquationsGravity}
    Q^{R}_{ab} + Q^{\varphi}_{ab} + g_{ab} Q = \tfrac{1}{2}T_{ab}
\end{equation}
where $T_{ab}$ is the stress tensor for matter degrees of freedom
(defined with respect to variations in $g_{ab}$), and
\begin{equation} \label{Eq:ELSTFieldEquationsGravityDefs}
\begin{aligned}
    Q^{R}_{ab} &= -2  c_{9}{} R_{ab}
                    -4 c_{2}{} R_{ac} R_{b}{}^{c}
                    - 4 c_{1}{} R_{ab} R
                    - 4 c_{3}{} R{_a}^{cde} R_{bcde} \\
                    &\quad
                    + 4 (c_{2}{} + 2 c_{3}{})
                    (R_{ac} R_{b}{}^{c} -  R^{cd} R_{a}{}_{cbd})  \\
                    &\quad
                    + 2 (2 c_{1}{} + c_{2}{} + 2 c_{3}{})
                    \nabla_{a}\nabla_{b}R
                    - 2 (c_{2}{} + 4 c_{3}{}) \Box R_{ab},\\
    Q^{\varphi}_{ab} &=
                    -2 c_{10}{} \varphi_{ab}
                    + 4(c_{5}{} - c_{8}) \varphi_{ab} \Box\varphi
                    + 4 c_{4}\varphi{_c}{_b} \varphi{^c}{_a} \\
                    &\quad
                    + 4(c_{7} + c_{8})(\varphi_{(a}\Box \varphi_{b)})
                    + 2(2 c_{4} + c_{5} - c_{8})
                      \varphi^{c} \nabla_{c}\varphi_{ab}\\
                    &\quad
                    - 2 c_{4} (R \varphi_{a} \varphi_{b}
                    + R_{ab} X - 2 R_{acbd} \varphi^{c} \varphi^{d})
                    - 4 c_{6}{} \varphi_{a}\varphi_{b} X \\
                    &\quad
                    -4\varphi^{c}(c_{5} + c_{7})R_{c(a} \varphi_{b)},\\
    Q &=    L-(c_5+2 c_7) (\Box\varphi)^2+(c_5+2 c_7)
                    R^{cd} \varphi_c \varphi_d  \\
                    &\quad
                    - (4 c_1+c_2) \Box R
                    - 2 (2 c_4+c_5+c_7) \varphi^c \Box \varphi_c \\
                    &\quad
                    - (4c_4+c_5) \varphi_{cd} \varphi^{cd}.
\end{aligned}
\end{equation}
Combined with $\nabla_a(J^a-\Phi^a)=0$, Eqs.
\eqref{Eq:ELSTFieldEquationsScalarCur}-\eqref{Eq:ELSTFieldEquationsGravityDefs}
yield the field equations for the action given in Eqs.
\eqref{Eq:ELSTAction} and \eqref{Eq:ELSTLagrangian}. These expressions
were obtained using the {\it xTras} add-on of the {\it xAct} package for
{\it Mathematica} \cite{xAct_web}.

\begin{figure}[]
    \includegraphics[width=0.88\columnwidth]{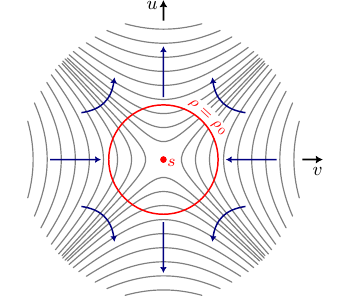}
    \caption{Contour plot for scalar field near a saddle point $s$ in
    the $u$-$v$ plane. The solid blue arrows indicate the direction of
    increasing scalar field $\varphi$, and the gray contours indicate
    level surfaces of the scalar field profile in Eq.
    \eqref{Eq:ScalarSaddleProfile}. The circle represents the surface of
    degenerate ${\bf{g}}_{\mu\nu}$ dividing regions where
    ${\bf{g}}_{\mu\nu}$ has Euclidean signature $\rho<\rho_0$ from
    Lorentzian signature regions $\rho>\rho_0$.}
    \label{fig:SaddleContours}
\end{figure}

%-----------------------------------------------------------------------
%   Microscopic model for SCDSs
%-----------------------------------------------------------------------
\subsection{Microscopic model for SCDSs}
We now ask whether the field equations admit regular solutions and
resemble SCDSs at large distances. Since the gradient of the scalar
field $\varphi_a$ corresponds to a timelike direction in the Lorentzian
phase, one can treat $\varphi$ as a time coordinate; from the direction
of time indicated in Fig. \ref{fig:QRSGluing}, one may infer that the
level surfaces of $\varphi$ for a regularized SCDS should resemble those
of a saddle, as illustrated in Fig. \ref{fig:SaddleContours}. If one
chooses $g_{\mu\nu}=\text{diag}(1,1,1,1)$ with coordinates $(u,v,y,z)$,
then a simple saddle profile for $\varphi$ is a harmonic one of the
form,
\begin{equation} \label{Eq:ScalarSaddleProfile}
    \varphi=(u^2-v^2)/(2 l_0),
\end{equation}
which for the flat metric satisfies $\Box \varphi=0$, with
$\varphi_{\mu\nu}$ being constant. The contours of $\varphi$ are
illustrated in Fig. \ref{fig:SaddleContours}. One finds that the Ricci
tensor for the effective metric ${\bf{g}}_{\mu\nu}$ vanishes for large
$\rho:=\sqrt{u^2+v^2}$, and that ${\bf{g}}_{\mu\nu}$ becomes degenerate
for $\rho=\rho_0:=\sqrt{X_C} l_0$, with Euclidean signature for
$\rho<\rho_0$ and Lorentzian signature for $\rho>\rho_0$. It should be
remarked that the singularity has only been regularized with regard to
the variables $g_{\mu\nu}$ and $\varphi$; the effective metric
${\bf{g}}_{\mu\nu}$ becomes singular at the degenerate surface
$\rho=\rho_0$, which we refer to as $s_d$. From the perspective of the
effective Lorentzian geometry, $s_d$ is a spacelike curvature
singularity (which follows from the vanishing of the determinant of
${\bf{g}}_{ab}$ as the surface $s_d$ is approached). Thus, we have
provided here a simple microscopic model for an SCDS in terms of regular
quantities; this is the sense in which an ELST regularizes an SCDS.

However, Eq. \eqref{Eq:ScalarSaddleProfile} is only an exact solution of
the field equations for flat $g_{\mu\nu}$ for the parameter choice,
\begin{equation} \label{Eq:ScalarSaddleProfileParameterSoln}
    c_{4}=c_{6}=c_{10}=0, \qquad \qquad l_0^2=(c_{5}-c_{8})/c_{11}.
\end{equation}
This parameter choice is potentially problematic, as one
might wish to have some dynamical mechanism that drives $X$ to a finite
value in the long-distance limit and a simple way to achieve this is by
adjusting the parameters $c_{6}$ and $c_{10}$ so that the corresponding
terms form an effective potential for $X$. Setting $c_{6}=c_{10}=0$
excludes this possibility. It is nonetheless important to note that the
analysis here at the very least provides a proof-of-concept that the
microscopic resolution of SCDSs is possible in the context of ELSTs.

If one wishes to avoid such parameter restrictions, it is still possible
to construct from Eq. \eqref{Eq:ScalarSaddleProfile} a class of
approximate solutions in the neighborhood of the origin for general
parameter choices using the Riemann normal coordinate expansion, in
which the metric takes the following form \cite{Muller:1997zk}:
\begin{equation} \label{Eq:MetricRiemannNormal}
    \begin{aligned}
    g_{\mu\nu} &= \delta_{\mu\nu}
                 - \frac{1}{3}\left[R_{\mu\alpha\nu\beta}\right]_0
                   x^\alpha x^\beta
                 - \frac{1}{6}\left[
                   \nabla_\gamma R_{\mu\alpha\nu\beta}\right]_0
                   x^\alpha x^\beta x^\gamma \\
               &\quad
                 - \left[\frac{2}{45}R_{\mu\alpha\lambda\beta}
                   R{^\lambda}_{\gamma\delta\nu}
                   +
                   \frac{1}{20}\nabla_\gamma \nabla_\delta
                   R_{\mu\alpha\nu\beta}
                   \right]_0
                   x^\alpha x^\beta x^\gamma x^\delta \\
                   &\quad
                 + O(x{^\cdot}^5),
    \end{aligned}
\end{equation}
where $[\cdot]_0$ indicates evaluation at the origin $x^\mu=0$. Upon
evaluating the field equations at the origin, one finds that for Eq.
\eqref{Eq:ScalarSaddleProfile} $[\Box\varphi_a]_0=[\varphi_a]_0=0$, 
and it follows that $[J^a]_0=0$. With the addition of an 
$O(x{^\cdot}^3)$ term to Eq. \eqref{Eq:ScalarSaddleProfile}, one can 
choose the corresponding coefficient to satisfy the field equation
$[\nabla_a(J^a-\Phi^a)]_0=0$, so that the scalar equations are
satisfied at the origin. From the gravity equations, the coefficients
have to satisfy
\begin{equation} \label{Eq:ELSTFieldEquationsGravityCoeffs}
    \begin{aligned}
        &   \biggl[4 \{(c_{2}{} + 4 c_{3}{}) \Box R_{\mu \nu}
            - (2 c_{1}{} + c_{2}{} + 2 c_{3}{})
                \nabla_{\mu}\nabla_{\nu}R\}
            + 4 c_{10}{} \varphi_{\mu \nu} \\
            &
            + 8 \{c_{3}{} R{_\mu}^{\alpha \beta \gamma}
                        R_{\nu \alpha \beta \gamma}
            - (c_{2}{} + 2 c_{3}{})
            (R_{\mu \alpha} R_{\nu}{}^{\alpha}
             - R^{\alpha \beta } R_{\mu}{}_{\alpha \nu \beta })\}\\
            &
            + 8 \{c_{2}{} R_{\mu \alpha} R_{\nu}{}^{\alpha}
                - c_{4}\varphi{_\alpha}{_\mu} \varphi{^\alpha}{_\nu}\}
            + 4(c_{9}{}+2c_{1}{} R) R_{\mu \nu} \\
                        &+
                        2 g_{\mu \nu}
                        \biggl\{(4 c_1+c_2) \Box R - c_{11} - c_9 R
                        - c_1 R^2
                        - c_2 R_{\alpha \beta} R^{\alpha \beta} \\
                        &\qquad
                        - c_3 R_{\alpha \beta \gamma \delta}
                              R^{\alpha \beta \gamma \delta}
                        - (c_8-4c_4-c_5)
                          \varphi_{\alpha \beta} \varphi^{\alpha \beta}
                        \biggr\}
            +T_{\mu \nu}\biggr]_0\\
            &
            =0.
    \end{aligned}
\end{equation}
Now since ${(\nabla \varphi)^2}/{X_C} \sim (x^\cdot)^2/\rho_0^2$, one
may require that the scale of the curvature satisfies
$\left[R_{\mu\alpha\nu\beta}\right]_0 \ll 1/\rho_0^2$ within some
neighborhood of the origin $x^\mu=0$, and in the same neighborhood, the
effective metric ${\bf{g}}_{\mu \nu}$ has Euclidean signature for $\rho
\lesssim \rho_0$ and Lorentzian signature for $\rho \gtrsim \rho_0$.
This can in principle be done in a neighborhood where
$O(x{^\cdot}^2)$ terms are dominant, since the derivatives
$\left[\Box R_{\mu \nu}\right]_0$ can be chosen independently of
$\left[R_{\mu\alpha\nu\beta}\right]_0$. Of course, the conditions under
which the solution continues to resemble an SCDS requires a more
complete solution, which will be considered elsewhere.

As we remarked earlier, if a theory admits regularized SCDS solutions,
it should also provide a mechanism for suppressing topological
structures such as those arising in Deutsch-Politzer and teleporter
spacetimes. Here, we briefly describe one possible mechanism for the
ELST we have considered here. The action defined in Eqs.
\eqref{Eq:ELSTAction} and \eqref{Eq:ELSTLagrangian}, being quadratic in
the curvature, contains a topological term, which yields the Euler
characteristic of the integration domain and would contribute a large
change in the value of the action for such structures compared to a
simply connected domain. Moreover, one might expect Deutsch-Politzer or
teleporter structures to lower the Euler characteristic (such structures
would, for instance, correspond to an increase in genus for a closed
manifold). Such topological configurations may be suppressed, provided
that the coefficient in front of the Euler characteristic in the action
is properly chosen and that the SCDS is regularized as prescribed above.

%=======================================================================

%-----------------------------------------------------------------------
%-----------------------------------
%-----------------
%--------
%---
%-
%
%
%-
%---
%--------
%-----------------
%-----------------------------------
%-----------------------------------------------------------------------

%=======================================================================
%-----------------------------------------------------------------------
%
%       Summary and Discussion
%
%-----------------------------------------------------------------------
%=======================================================================
\section{Summary and Discussion}\label{Sec:Disc}
\begin{figure}[]
    \includegraphics[width=1.0\columnwidth]{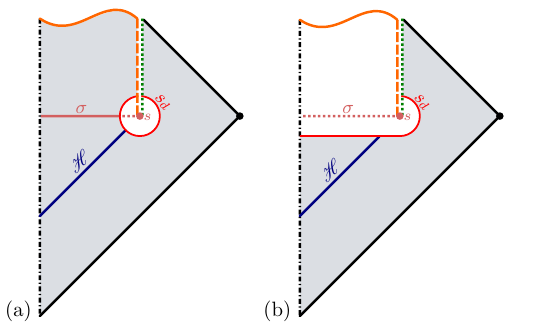}
    \caption{Possible regularization scenarios for an evaporating black
    hole in an ELST. As in Figs. \ref{fig:BHevapTrousers} and
    \ref{fig:RContours}, $\mathscr{H}$ is the horizon, and $\sigma$ and
    $s$ indicate where the respective Schwarzschild and end point
    singularities would otherwise be. The effective metric
    ${\bf{g}}_{ab}$ is Lorentzian in the shaded regions, and the red
    contour $s_d$ is the surface where ${\bf{g}}_{ab}$ becomes
    degenerate.}
    \label{fig:RegPoss}
\end{figure}
In this article, we considered the end point of an evaporating $1+1$
black hole. We have shown how given a theory that regularizes the
Schwarzschild singularity in the Lorentzian regime, the end state of an
evaporating $1+1$ black hole can be characterized by a particular type
of quasiregular singularity, which we have termed a saddlelike causally
discontinuous singularity (SCDS). We have described how such
singularities can be described in terms of regular quantities within the
framework of emergent Lorentz signature theories (ELSTs)
\cite{Mukohyama:2013ew,Kehayias:2014uta}, in which the notion of time
and Lorentz signature emerge from a Euclidean scalar-tensor theory. In
particular, we consider the Euclidean signature quadratic curvature
scalar-tensor theory of \cite{Mukohyama:2013gra}, which evades the
Ostrogradsky instability (as the theory is fundamentally Euclidean) and
is perturbatively renormalizable \cite{Muneyuki:2013aba}. In particular,
we showed that the classical field equations for the purely
gravitational part of the aforementioned ELST admits an exact solution
corresponding to a regularization of SCDSs for a particular choice of
parameters, and for more general parameter choices, a class of
approximate solutions in the neighborhood of a saddle point in the
scalar field profile can be obtained. These results provide a
proof-of-concept that the microscopic resolution of SCDSs is possible in
the context of ELSTs. In addition to their utility for describing the
end point of black hole evaporation and topological structures such as
Deutsch-Politzer and teleporter spacetimes, the solutions we have
presented may be of use for regularizing other types of naked
singularities \cite{Vaz:1993eg,Harada:2000me}.

Of course, there are plenty of issues to resolve before our main
proposals, namely that the the solutions we have found describe SCDSs at
long-distance scales, and that the regularized SCDS solutions describe
the end state of an evaporating black hole, can be conclusively
established. A resolution of the former will ultimately require a more
complete numerical or analytical effort; an appropriate starting point,
perhaps, would be an exploration of the solution space neighborhood
about the solutions we have considered---in particular, one might begin
by performing a linear perturbation analysis (accompanied by an
investigation into questions of linearization stability, the matter of
whether solutions of the linearized equations are in fact tangent to the
solution space for the nonlinear equations).

Whether the field equations of an ELST does indeed regularize the
singularities of an evaporating spherically symmetric \footnote{One might
expect a rotating black hole undergoing an extended period of Hawking
radiation to lose its angular momentum \cite{Page:1976ki}, so one can
argue that the restriction to spherically symmetric black holes is
appropriate. Of course, black holes with a history of rotation (and
charge) may have a radically different singularity structure in the
interior, but one might expect the regularization to yield qualitatively
similar results if the Euclidean region is restricted to a compact
neighborhood of the end point.} black hole will also require additional
effort. We have only shown that this is a possibility in ELSTs, but a
more complete analysis is required to conclusively demonstrate this. The
solutions presented in Sec. \ref{Sec:ELSTs} assume the Schwarzschild
singularity is regularized so that the effective metric is
${\bf{g}}_{ab}$ only has Euclidean signature in a compact neighborhood
of the end point $s$ of the black hole; this scenario is illustrated in
Fig. \ref{fig:RegPoss} (a); ${\bf{g}}_{ab}$ is Lorentzian in the shaded
region. However, another possibility, illustrated in Fig.
\ref{fig:RegPoss} (b), is that the regularization of both the end point
$s$ and Schwarzschild singularity occur in a region where
${\bf{g}}_{ab}$ has Euclidean signature beyond the boundary $s_d$ of
region where ${\bf{g}}_{ab}$ has Lorentzian signature. One can easily
imagine plenty of other regularization scenarios; for instance, one can
imagine ``islands'' or noncompact regions in Fig. \ref{fig:RegPoss} (b)
below the surface $s_d$ in which the effective metric ${\bf{g}}_{ab}$
has Lorentzian signature, or ``islands'' of Euclidean signature in Fig.
\ref{fig:RegPoss} (a). In either scenario, such a regularization would
yield a ``baby universe'' resolution
\cite{Hawking:1988ae,Hawking:1988wm,Hawking:1990jb} to the famed black
hole information paradox \cite{Hawking:1976ra} (see also
\cite{Preskill:1992tc,*Hossenfelder:2009xq} for a discussion of possible
resolutions), though the resulting universe need not admit a Lorentz
signature metric.

We have only considered one possible scenario in which a candidate
ultraviolet completion for quantum gravity can regularize SCDSs (though
we emphasize that our analysis is entirely classical). Various
approaches to signature change have been extensively explored in the
literature \cite{Ellis:1991st,Ellis:1991sp,Dray:1991zz,Greensite:1992xp,
Greensite:1992np,Hayward:1992zp,Carlini:1993up,Mars:1993mj,
Kerner:1993fm,Dereli:1993pj,Dray:1993xu,Hayward:1994ik,
Martin:1994rs,Hellaby:1994ww,Kriele:1994ci,Greensite:1994sj,
Carlini:1995gn,Embacher:1995xp,Dray:1995fb,Embacher:1995wt,
Martin:1995cx,Altshuler:1996ae,BarberoG:1995tgc,Dray:1996cx,
Dray:1996dc,Dray:1996cw,Iliev:1998sv,Dray:2000hb,Mars:2000gu,
Wetterich:2004za,Darabi:2004bh,Borowiec:2007zz,Pedram:2008sj,
White:2008xr,Vakili:2013fra,Magueijo:2013yya,Mielczarek:2014kea,
Ambjorn:2015qja,Barrau:2016sqp,Nissinen:2017akm,Zhang:2019rrc,
Bondarenko:2021xvz,Bondarenko:2022krf}, and from time to time, it has
been suggested, based on quantum gravitational considerations, that the
metric should be complex valued
\cite{Louko:1995jw,Ivashchuk:1987mnk,Halliwell:1989dy,Lyons:1992ua,
Hayward:1995bi,Sorkin:2009ka,Visser:2017atf}; allowability criteria for
complex metrics have recently become a topic of interest
\cite{Kontsevich:2021dmb,Witten:2021nzp,Lehners:2021mah,Visser:2021ucg,
Loges:2022nuw,Briscese:2022evf,Jonas:2022uqb,Jonas:2023hcl}. Another
possible avenue to consider may be a generalization or modification of
the Callan–Giddings–Harvey–Strominger model \cite{Callan:1992rs} or
related toy models \cite{Russo:1992ax,Lowe:1993} to incorporate
signature change or some emergent Lorentz signature mechanism. It may be
of interest to consider the regularization of SCDSs in these alternative
approaches, as any quantitative or qualitative differences between the
different approaches may eventually lead to distinct predictions for any
signals from the final stages of black hole evaporation.

Nevertheless, for the scenarios in which the end state of an evaporating
black hole is an apparent naked singularity or a region of flat
spacetime, the approach based on the ELST we have considered in this
paper is of general interest for two reasons. First, the ELST of
\cite{Mukohyama:2013gra} can be regarded as the perturbatively
renormalizable part for a wide class of effective field theories, so it
may be relevant even if the ultimate UV completion takes a radically
different form. Second, the solutions we have presented are {\it
classical} and may either provide a starting point for a saddle point
analysis in the full quantum theory or a semiclassical description for a
possible end state of an evaporating black hole.

%=======================================================================

%-----------------------------------------------------------------------
%-----------------------------------
%-----------------
%--------
%---
%-
%
%
%-
%---
%--------
%-----------------
%-----------------------------------
%-----------------------------------------------------------------------

%=======================================================================
%		ACKNOWLEDGMENTS
%=======================================================================

\begin{acknowledgments}
\it J.C.F. thanks R. A. Matzner, R. Casadio, A. Kamenshchik, and P. Chen
for comments and feedback, and is also grateful to Charles University in
Prague, Istituto Nazionale Di Fisica Nucleare (INFN) - Sezione di
Bologna, CENTRA, Instituto Superior T\'ecnico, University of Lisbon, and
the Center for Gravitational Physics at The University of Texas at
Austin for hosting visits during which part of this work was conducted.
J.C.F. also acknowledges support from the Leung Center for Cosmology and
Particle Astrophysics (LeCosPA), National Taiwan University (NTU), the
NTU Physics Department, and the R.O.C. (Taiwan) National Science and
Technology Council (NSTC) Grant No. 112-2811-M-002-132. The work of S.M.
was supported in part by the World Premier International Research Center
Initiative (WPI), MEXT, Japan. The work of S.C. has been carried out in
the framework of activities of the INFN Research Project QGSKY.
\end{acknowledgments}

%=======================================================================
%		BIBLIOGRAPHY
%=======================================================================

\bibliography{ref}

\end{document}